\documentstyle[12pt,epsf,rotate]{article}

\begin{document}
\
\vspace{8\baselineskip}

\noindent{\large \bf NETWORKS OF STEPS ON AU AND PT CRYSTALS}

\vspace{2 \baselineskip}

\newenvironment{add}
{\begin{list}{}{\setlength{\leftmargin}{1truein}}
      \item[]}
{\end{list}}
\begin{add}
{\bf Henk van Beijeren}\\
\\
Institute for Theoretical Physics\\
Universiteit Utrecht\\
3508-TA Utrecht, The Netherlands\\
\\ \\
{\bf Enrico Carlon}\\  \\
Institute for Theoretical Physics\\
Katholieke Universiteit Leuven\\ 
B-3001 Leuven, Belgium
\end{add}

\vspace{2\baselineskip}

\begin{abstract}
\noindent
Networks of steps, seen in STM observations 
of vicinal surfaces on Au and Pt (110), are analyzed.
A simple model is introduced for the calculation of the free 
energy of the networks as function of the slope parameters,
valid at low step densities.
It predicts that the networks are unstable, or at least metastable, 
against faceting and gives an equilibrium crystal shape with sharp 
edges either between the (110) facet and rounded regions or between 
two rounded regions.
Experimental observations of the equilibrium shapes of Au or Pt
crystals at sufficiently low temperatures, i.e. below the
deconstruction temperature of the (110) facet, could 
check the validity of these predictions.
\end{abstract}
\vspace{\baselineskip}
\noindent{\bf INTRODUCTION}
\vspace{\baselineskip}

(110) surfaces of fcc metals have been 
intensively studied in the past years by means of several
experimental techniques. 
It was found by scattering experiments, and subsequently
confirmed by direct STM observations that some ``heavy" metals
such as Au, Pt, Ir form a $(2 \times 1)$ reconstructed state.
Other metals, e.g. Pb and Al, are not reconstructed.

The $(2 \times 1)$ structure is also known as missing--row reconstruction
because one out of two rows of atoms, aligned along the $[1 \bar{1}0]$
direction,
is missing from the surface layer. We stress also that there
are two different 
realizations of these reconstructed states, 
in which either the even
or the odd rows are missing.

\begin{figure}[t]
\epsfxsize=8cm
\centerline{\rotate[r]{\epsfbox{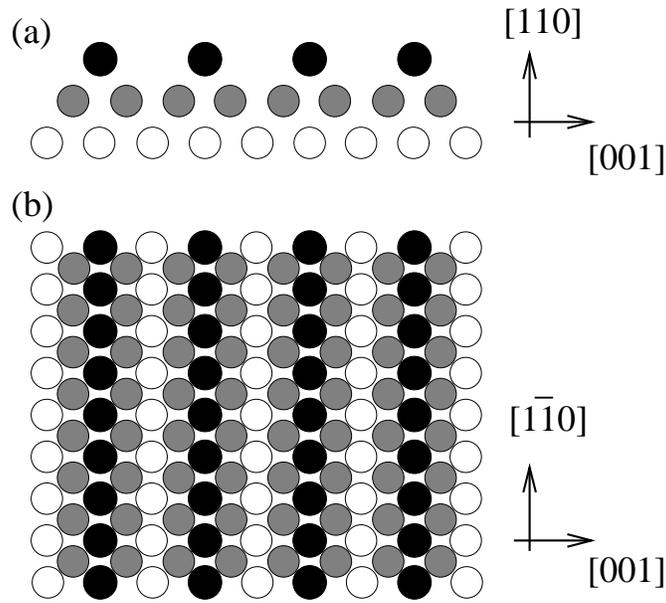}}}
\caption{(a) Profile and (b) top view of a missing row reconstructed
 (110) surface.}
\label{FIG01}
\end{figure}

Fig.\ \ref{FIG01} shows a top and a side view of the missing--row reconstructed 
structure. Along the 
%%EE $[100]$ 
$[001]$
%%EE  IT IS BETTER TO USE THE SAME CONVENTION OF THE FIGURE 1
direction the surface assumes a hill-and-valley
profile, where the sides of the hills are actually $(111)$ microfacets.
In the $(111)$ orientation, surface atoms are closely packed, therefore 
such orientations are energetically favored.

\begin{figure}[h]
\centerline{\epsfbox{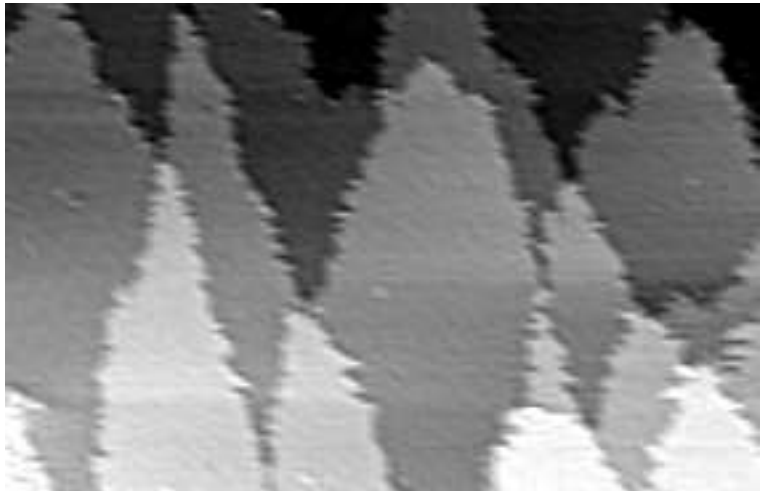}}
\caption{STM images of vicinals of $Au(110)$ (courtesy of
 M.S.~Hoogeman, L.~Kuipers and J.W.M.~Frenken, AMOLF Amsterdam). 
 The area shown is of $190 nm \times 120 nm$ at $T=550 K$, 
 with a miscut angle of $0.07$ degrees.}
\label{FIG02}
\end{figure}

In several STM investigations (Gritsch et al.\, 1991; Gimzewski et al.\, 
1992; Kuipers, 1994) of vicinal 
orientations of (110) Au and Pt surfaces an unusual pattern 
of steps, as shown in Fig.\ \ref{FIG02}, was observed. 
The figure represents a surface orientation slightly tilted 
towards the $[1 \bar{1} 0]$ direction with respect to the $(110)$ 
facet.
The missing rows, which are not visible in the figure, run along 
the vertical direction.

This pattern of steps is unusual indeed. In normal 
situations a miscut
along the $[1 \bar{1} 0]$ direction is generated by steps that run
perpendicular to the missing rows, i.e. along the horizontal
direction of Fig.\ \ref{FIG02},
because such an arrangement minimizes the total length and hence the total free
energy of the steps required to produce the miscut.
Instead, on missing-row reconstructed surfaces the steps zig-zag and repeatedly 
touch each other at a 
collection of contact points; 
one can also say that they form a {\em network} of two arrays of 
roughly parallel steps crossing each other and forming on average
angles $\phi$ and $-\phi$, with the vertical direction in Fig.\ \ref{FIG02}.
An explanation for the formation of this network has been given
already in (Kuipers, 1994). In this article we give a more quantitative description
based on a simple, yet, we think, quite realistic model,
which describes surface orientations close to the $(110)$ facet.
On the basis of this model we discuss the thermodynamical properties of
the networks (Carlon and Van Beijeren, 1996). 
We find that the network is actually unstable
or at least metastable:
in equilibrium 
it decays into a combination of stable surface orientations, with 
the appearance of sharp edges in 
between. As a consequence the edge of the $(110)$ facet shows cusps,
i.e. jumps in the direction of its tangent.

%%EE ACTUALLY THERE IS NO SECTION NUMBERING ANYMORE, SO I WOULD LEAVE
%%EE  OUT THE NEXT SENTENCES
%%EE
%%EE This article is organized as follows: in Sec. 2 we
%%EE introduce the model and calculate the surface free energy for
%%EE orientations close to the $(110)$ facet, in Sec.\ \ref{sec:ecs}
%%EE we present the resulting equilibrium shape
%%EE and discuss the instabilities occurring in this free energy. 
%%EE
\vspace{2\baselineskip}
\noindent{\bf THE MODEL}
\label{sec:mod}
\vspace{1\baselineskip}

\begin{figure}
\centerline{\epsfxsize=5cm \rotate[r]{\epsfbox{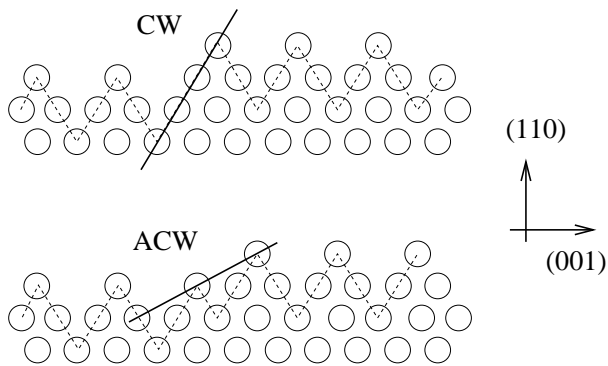}}}
\caption{Examples of clockwise (CW) and anticlockwise (ACW) steps.}
\label{FIG03}
\end{figure}

In a missing--row reconstructed (110) facet one can distinguish 
two different types of steps parallel to the missing rows, 
commonly known as clockwise and anticlockwise steps and 
illustrated in Fig.\ \ref{FIG03}. 
%%EE TOO MANY OBSERVATIONS ....
In STM observations 
anticlokwise steps are rarely 
%%EE observed, 
seen
%%EE
especially at low
temperatures, and when 
%%EE observed 
present
%%EE
they are mostly pinned by impurities.
This indicates that clockwise steps have markably lower free energies
per unit of length than the anticlockwise ones; this energy difference
can in principle be derived from STM observations, by estimating the relative
frequencies at which step segments of given length of either type of step do
occur.

On the basis of STM observations on Au$(110)$ surfaces it has been 
concluded (Kuipers, 1994) that at room temperature and for 
sufficiently clean samples, anticlockwise steps 
should be absent from the surface.

\begin{figure}
\centerline{\epsfxsize=8cm \rotate[r]{\epsfbox{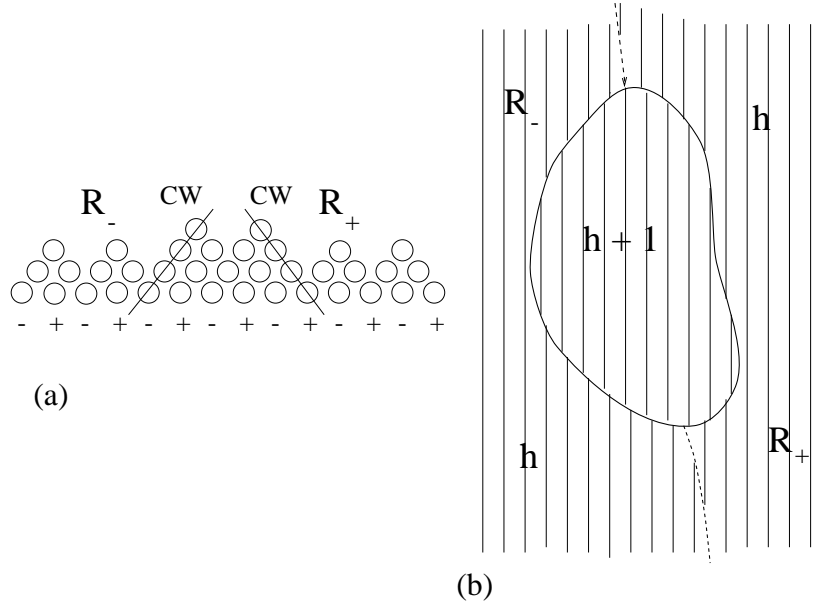}}}
\caption{(a) A clockwise step up followed by a clockwise step
 down generate a shift in the reconstruction order; $R_-$ and $R_+$
 indicate the two possible reconstructed phases.
 (b) As a consequence of this a closed clockwise step must be accompanied
 by domain boundaries between the two phases $R_-$ and $R_+$.}
\label{FIG04}
\end{figure}

Due to the presence of reconstruction not all possible configurations of
clockwise steps will actually occur on the surface. 
As illustrated in Fig.\ \ref{FIG04}(a) 
a clockwise step going up followed by another clockwise 
step going down induces a shift in the reconstruction
of the lower terrace. For this reason, a 
closed clockwise step cannot be formed 
on a given terrace unless it is accompanied by domain boundaries 
separating regions of different reconstruction order (Fig.\ \ref{FIG04}(b)). 

This fact may have some important consequences, for instance, on how 
the $(110)$ facet roughens. In the present work however we will 
focus on the behavior of the system far from the deconstruction or roughening
critical points.

The interplay between step orientation and surface reconstruction
is essential for understanding the formation of the network of steps.
Indeed, a step perpendicular to the missing rows cannot 
zig--zag forming clockwise segments parallel to the missing rows, if
expensive domain boundaries between opposite reconstruction states
are to be avoided. 

\vspace{1\baselineskip}
\noindent{\bf The single step free energy}
\vspace{1\baselineskip}

To clarify and quantify the previous discussion we calculate the free energy of
an isolated step as function of the step orientation.
For convenience we consider, for the energy of an anticlockwise
step segment, the limit $E_{ACW} \to \infty$: only clockwise step
segments are allowed and they have energies per unit 
lengths $\delta_y$ and $\delta_x$ where $y$ and $x$ 
are the directions parallel respectively perpendicular to the missing
rows.
One could eventually introduce more parameters, such
as a corner  energy, but this is not essential.

\begin{figure}
\centerline{\epsfxsize=6cm \rotate[r]{\epsfbox{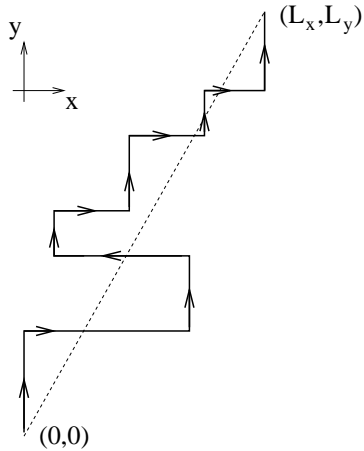}}}
\caption{Example of microscopic configuration of a step; if anticlockwise
 steps are to be avoided only segments along the $+y$ and $\pm x$ directions
 are allowed.}
\label{FIG05}
\end{figure}

Let us consider the step as an oriented walk from a lattice point
$(0,0)$ to $(L_x,L_y)$, as shown in Fig.\ \ref{FIG05}.
As seen above, if anticlockwise steps are to be avoided, only step
segments in the $\pm x$-directions and the $+y$-direction are allowed.
A simple calculation (for more details see Appendix\ I) gives 
the following free energy of a step per unit of length, tilted over
an angle $\phi$ with respect to the missing rows:

\noindent
\begin{eqnarray}
f_s (\phi) = \frac{\ln z(\phi)}{2 \beta} \,\,\,|\sin\phi| +
\left[\delta_y + \frac{\sqrt{2}}{\beta} 
\ln \left( \frac{2 \cosh(2 \beta \delta_x)-(z(\phi)+1/z(\phi))}
{2 \sinh(2 \beta \delta_x)} \right )\right] \cos\phi
\label{singst}
\end{eqnarray}
where $\beta = 1/(k_B T)$ and:
\begin{eqnarray}
z(\phi) &=& \frac{\cosh(2 \beta \delta_x) \, t(\phi)+
\sqrt{1+\sinh^2(2 \beta \delta_x) \, t^2(\phi)}}{1+t(\phi)}
\end{eqnarray}
with $t(\phi) = |\tan\phi|/2\sqrt{2}$.
Notice that the limiting value for the free energy of a step 
running perpendicular to the missing
rows is simply $f_s(\pi/2) =
\delta_x$: such steps are perfectly straight and their
free energy is ``frozen", i.e. temperature independent. 
In the limit $\phi \to 0$, where steps are parallel to the 
missing rows, the step free energy is:
\begin{eqnarray}
f_s(0) &=& \delta_y + \frac{\sqrt 2}{\beta} \ln 
\left(\tanh( \beta \delta_x) \right)
\end{eqnarray}
where the first term on the r.h.s. is the energy of the ground state
configuration, while the second term is negative and represents the 
contribution of thermal excursions along the $x$-direction. 

\begin{figure}
\centerline{\epsfysize=6cm \epsfbox{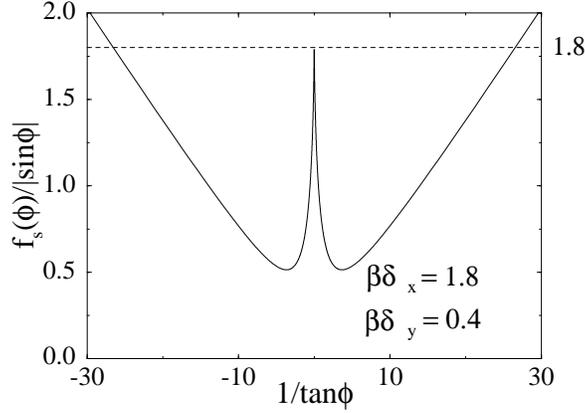}}
\caption{The step free energy per unit of projected length along the
 $x$-direction as function of $1/\tan \phi$. Step with orientations
 close to $\phi = \pi/2$ are unstable as can be seen from the cusp in
 the figure.}
\label{FIG06}
\end{figure}

For investigating the stability of steps of various orientations
it is convenient to calculate $f_s(\phi)/|\sin\phi|$, 
the step free energy per unit of projected length along the
$x$-direction (Van Beijeren and Nolden, 1987). 
This quantity is shown in Fig.\ \ref{FIG06}: it has a local 
maximum with a cusp at $\phi=\pi/2$. For steps slightly inclined 
with respect to this orientation it decreases, since such steps 
have a larger entropy.
This can be seen expanding $f_s(\phi)/|\sin\phi|$ around 
$\phi = \pi/2 \pm \epsilon$; such an expansion yields:
\begin{eqnarray}
\left.\frac{f_s(\phi)}{|\sin \phi|} \right|_{\frac{\pi}{2} \pm \epsilon}
&=& f_s(\pi/2) + A |\epsilon| - B |\epsilon \log \epsilon| + \ldots
\end{eqnarray}
with $A$ and $B$ some non-negative constants. The second term on the r.h.s.
is due to the energy of segments of steps parallel to the $y$-direction, 
while the third term is the decrease in free energy due to entropy. The 
entropic term dominates at sufficiently small $|\epsilon|$, except at $T=0$, 
where $B=0$.

Obviously $f_s(\phi)/|\sin\phi|$ diverges for $\phi \to 0$. 
It is minimal for an angle $\phi_0$ satisfying\footnote{The quantity under 
the square root in (\ref{angle}) becomes negative at small 
values of $\beta \delta_x$ and $\beta \delta_y$; this happens at temperatures 
above the roughening temperature of the $(110)$ surface when the 
solid-on-solid approximation for the step free energy $f_s(\phi)$ is not 
positive definite and the simple theory considered here breaks down.}:
\begin{eqnarray}
t(\phi_0) &=& 
\frac{\sqrt{\left( \cosh(2 \beta \delta_x) - e^{-\beta
\delta_y/\sqrt{2}} \sinh(2 \beta \delta_x) \right)^2 - 1 }}
{e^{- \beta \delta_y/\sqrt{2}}\sinh(2 \beta \delta_x)}.
\label{angle}
\end{eqnarray}

This simple calculation shows that, on a surface tilted along the 
%%EE $[\bar{1}10]$ 
%%EE BETTER USE ALWAYS THE SAME NOTATION
$[1\bar{1}0]$ 
%%EE
direction, steps making a finite angle $\phi_0 \neq \pi/2$
with respect to the missing rows are entropically favored.
$\phi_0$ is the optimal value of the angle of the network.
Its value, given by Eq.\ (\ref{angle}), could in principle be used to determine 
the parameters $\delta_x$ and $\delta_y$ from measured values of 
$\phi_0$ at different temperatures. 
In fact the temperature dependence of $\phi_0$
has been studied in some STM observations of networks on Au(110) 
surfaces by Hoogeman and Frenken (Hoogeman and Frenken, private communication). 
They found a characteristic angle that is rather temperature independent and 
pointed
out this may be due to the presence of impurities which act as 
pinning sites for the network. We think that this point is quite
interesting and deserves further experimental investigations.

\vspace{1\baselineskip}
\noindent{\bf The equilibrium shape of the $(100)$ facet}
\vspace{1\baselineskip}

\begin{figure}
\centerline{\epsfysize=7cm \epsfbox{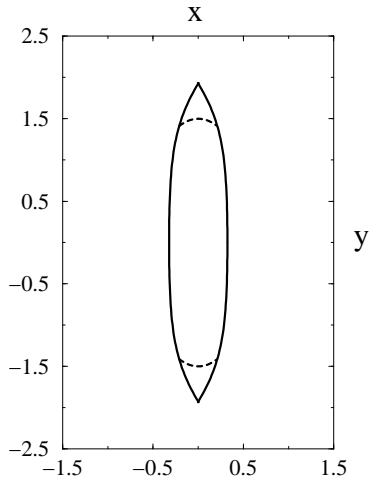}}
\caption{Equilibrium shape of the $(110)$ facet obtained from the Wulff
  construction, for the same values of $\beta \delta_x$ and $\beta \delta_y$
  as shown in Fig.\ \protect{\ref{FIG06}}. Notice the cusps, due to instabilities 
  of steps perpendicular to the missing rows.}
\label{FIG07}
\end{figure}

The equilibrium shape of the $(110)$ facet can be found 
by applying a one-dimensional Wulff construction 
to the step free energy as function of orientation (Van Beijeren and Nolden,
1987).
The result of this construction, for a representative choice of step energies 
and temperature, is shown in Fig.\ \ref{FIG07}. Steps with orientations close
to $\phi = \pi/2$ are unstable and would phase separate in combinations
of two steps of orientations 
$\phi_0$ and $\pi -\phi_0$:
the shape of the $(110)$ facet resembles that of an almond, with cusps
along the $[1 \bar{1} 0]$ direction.

This type of shape is unusual, because the cusps correpond
to first order phase transitions in a one-dimensional system at
finite temperatures. In normal situations this 
could not occur (Van Beijeren and Nolden, 1987):
if steps of orientations $\phi_0$
and $-\phi_0$
both have free energy, say $f_0$, 
steps of intermediate orientations could always be built out of an alternating
sequence of segments of orientations $\phi_0$ and $-\phi_0$. Then at any non-zero
temperature the lowering of free energy through entropy gain outwins
the increase through the excess energy at the corners between the successive 
segments, provided the density of these corners is sufficiently low.
This is a variant of the usual argument against phase transitions in 
one-dimensional systems at finite temperatures.
In the present case this argument does not work since, due to
topological constraints induced by the reconstruction, a 
configuration of short ``zig--zagging" step segments of 
orientations $\phi_0$ and $-\phi_0$ joined together cannot be 
formed, as pointed out above.

\vspace{1\baselineskip}
\noindent{\bf The surface free energy }
\vspace{1\baselineskip}

Next we calculate the surface free energy as function of the slope parameters 
$p$ and $q$, which denote the tangents of the tilting angles of the surface 
with the $[1\bar{1}0]$ and the $[001]$ directions. 
Both parameters $q$ and $p$ will be considered small, so we restrict
ourselves to vicinal surface orientations.

Interactions between steps are introduced in a simple way: a free 
energy $\tilde{\epsilon}$ is associated to each pair of crossing 
steps. $\tilde{\epsilon}$ keeps account of both short and long range 
interactions between the steps and it is defined as the excess free 
energy associated to the intersection point of two crossing steps 
forming angles $\phi_1$ and $\phi_2$ with the $y$-direction 
on a $(110)$ facet:
\begin{eqnarray}
\tilde{\epsilon}(\phi_1,\phi_2)  &=& F(\phi_1,\phi_2) 
- A f_0 - s_1 f_s(\phi_1) - s_2 f_s(\phi_2)
\label{crossfree}
\end{eqnarray}
Here $F(\phi_1,\phi_2)$ is the free energy of the surface of area $A$
with two crossing steps of lengths $s_1$ and $s_2$, $f_0$ is the free 
energy of the $(110)$ facet per unit of area (no steps are present), 
and $f_s(\phi)$ the free energy of a single step, whose properties 
have been discussed above.

\begin{figure}
\centerline{\epsfysize=9cm \epsfbox{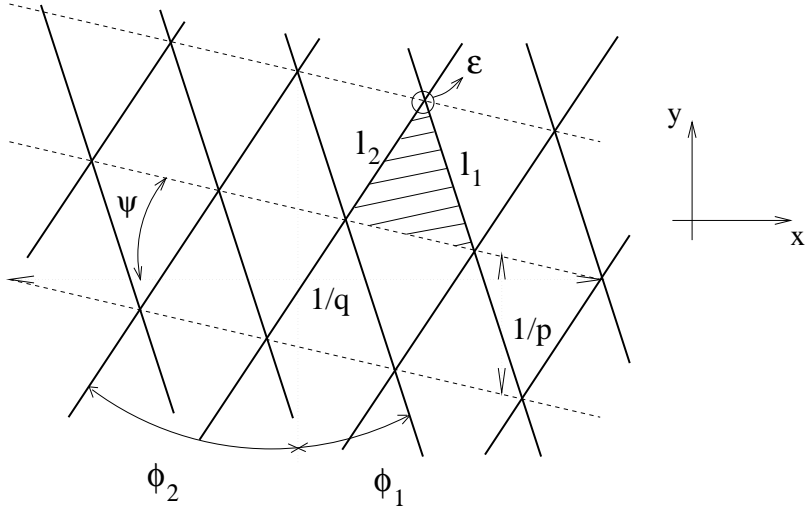}}
\caption{Schematic view of the network of crossing steps.}
\label{FIG08}
\end{figure}

One could try to estimate $\tilde{\epsilon}$ from a microscopic point 
of view taking into account long range interactions of elastic or entropic
type and short range interactions near the crossing point. This estimate
is non-trivial and for the general features of the equlibrium crystal
shape that we want to address here, only the sign of $\tilde{\epsilon}$ 
matters.
In general, it is expected (Bernasconi and Tosatti, 1993) that elastic interactions  
between two antiparallel steps (as are the crossing steps of the network,
provided $\phi_0$ is not too large) yield a negative contribution to 
$\tilde{\epsilon}$. 
Entropic interactions (Gruber and Mullins, 1967; Jayaprakash et al., 1984), 
which become relevant at 
higher temperatures, are repulsive and give a positive contribution
to $\tilde{\epsilon}$. Finally at the crossing point one would expect 
a positive contribution to $\tilde{\epsilon}$.
In the rest of the article we discuss both possibilities for the sign 
of $\tilde{\epsilon}$, which yield two different scenarios for 
the equilibrium shape of the crystal.

Let us consider a miscut along the $[1 \bar{1} 0]$ direction, generated by 
a pattern of crossing steps forming angles $\phi_1$ and $\phi_2$ with the 
$y$-axis, as shown in Fig.\ \ref{FIG08}. The dashed inclined lines indicate hypothetical 
parallel isolated steps that would generate the same macroscopic orientation 
as the network; $1/q$ and $1/p$ are the average distances between these steps 
along the $x$ and the $y$ direction.

The free energy per unit of projected area can be written as: 
\noindent\
\begin{eqnarray}
\tilde{f}(p,q,\phi_1,\phi_2) &=& f_0 + \frac{1}{2 A} \left[ l_1 f_s(\phi_1) 
\,+\, l_2 f_s(\phi_2) \,+\, \tilde{\epsilon}(\phi_1,\phi_2) \right]
\label{fretomin}
\end{eqnarray}
\noindent
where $f_0$ is the free energy per unit area of the $(110)$ facet, while 
$l_1$ and $l_2$ are the lengths of the two sides of the dashed triangle, 
of area $A$, shown in Fig.\ \ref{FIG08}. 

The actual free energy can be found by minimizing the free energy
(\ref{fretomin}) with respect to the angles $\phi_1$ and $\phi_2$; 
to lowest orders in $p$ and $q$ and for $\psi < \pi/2 - \phi_0$, 
this amounts to minimizing $f_s(\phi)/|\sin \phi|$.
Therefore the minimum of (\ref{fretomin}) is at $\phi_1 \approx
-\phi_2 = \phi_0 + O(p,q)$, where $\phi_0$ is the angle given by
(\ref{angle}).
Substituting this back into (\ref{fretomin}), expressing $A$, $l_1$, 
$l_2$ as function of $p$, $q$, $\phi_1$, $\phi_2$ and expanding to 
lowest orders in $p$ and $q$ we find:
\begin{eqnarray}
f(p,q) &=& f_0 + \frac{f_s(\phi_0)}{\sin\phi_0}\,p +
\frac{\tilde{\epsilon}}{2} \left(\frac{p^2}{\tan\phi_0}-\tan\phi_0 
\,q^2 \right) 
\label{fremin}
\end{eqnarray}
where $\tilde{\epsilon}$ is the interaction free energy (\ref{crossfree})
calculated at angles $\phi_1 = -\phi_2 = \phi_0$.
As in usual expansions of surface free energies around facets (see e.g.\
Van Beijeren and Nolden, 1987) the term linear in $p$ represents the contribution of 
non-interacting steps.
The interaction terms are quadratic in the step densities,
differently from usual step-step interactions which lead to
terms cubic in the step densities (Gruber and Mullins, 1967).
The origin of the quadratic term can be understood easily:
the number of step crossings per unit area is simply proportional
to the product of the densities, $p \pm q \tan \phi_0$, of
the two types of steps.
In addition there are long range interactions between the parallel 
steps in the network, but they will only contribute to terms of
cubic or higher order in the step density expansion of the free
energy (Gruber and Mullins, 1967).

For $\psi > \pi/2 - \phi_0$ the expression (\ref{fretomin}) is minimized 
by a single array of steps (so $\phi_1=\phi$ and $l_2=0$) and the free
energy takes the usual form (Gruber and Mullins, 1967):
\begin{eqnarray}
f(p,q) &=& f_0 + f_s(\phi)\,\sqrt{p^2+q^2} + 
O((p^2 + q^2)^{\frac{3}{2}})
\label{fusual}
\end{eqnarray}

Notice that the expression (\ref{fremin}) for the free
energy of the step network, irrespectively of the sign of
$\tilde\epsilon$, is a non-convex function of $p$ and $q$.
This result implies that the network is always unstable:
some surface orientations disappear from the equilibrium
shape of the crystal and are replaced by sharp
edges.

\vspace{2\baselineskip}
\noindent{\bf
EQUILIBRIUM CRYSTAL SHAPES
}
\vspace{1\baselineskip}
\label{sec:ecs}

As is well known (see, for instance, 
Van Beijeren and Nolden, 1987), the
equilibrium crystal shape is the shape that minimizes the
total surface free energy at a given fixed volume.
{F}rom the minimization of the free energy calculated above we 
can construct the equilibrium shape of the crystal around the
$(110)$ facet. This shape depends crucially on the sign of the
interaction free energy $\tilde{\epsilon}$, therefore in the
rest of the paper we will distinguish two different cases.

\vspace{1\baselineskip}
\noindent{\boldmath{$ \tilde{\epsilon} > 0$}}
\label{3 b}
\vspace{1\baselineskip}

\begin{figure}
\centerline{\epsfysize=5.5cm \epsfbox{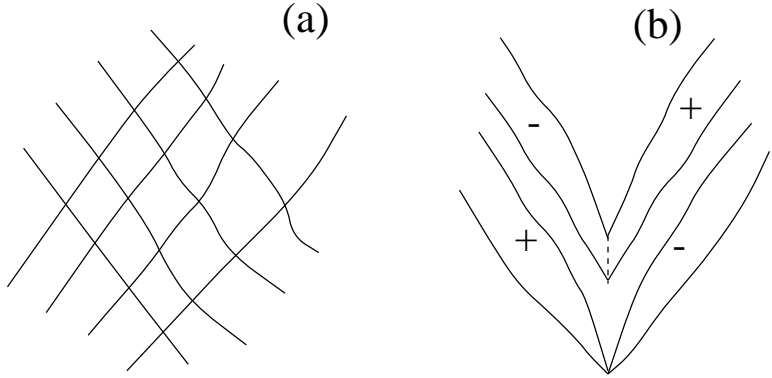}}
\caption{(a) The network of crossing steps, and (b) 
 a combination of two step arrays of different orientation, 
 after the phase separation, predicted by (5)
 for $\tilde{\epsilon} > 0$, has occurred. The $+$ and $-$ indicate 
 the opposite reconstruction phases induced by the clockwise steps.}
\label{FIG09}
\end{figure}

When $\tilde\epsilon$ is positive the system can decrease the 
surface free energy reducing the density of crossings.
By applying the Maxwell construction to Eqs. (\ref{fremin},
\ref{fusual}) one finds that a given surface orientation $(p,0)$
separates into two orientations of slopes $(p,p \tan\phi_0)$ and 
$(p,-p \tan\phi_0)$ joining at a ridge, as depicted in Fig.\ \ref{FIG09}(b).
In this combination of two surface orientations all the crossing 
points have been "eliminated'' with a net gain of free energy.
To generate the interface between the two surface orientations 
one has to pay a positive amount of boundary free energy, but 
this free energy will be proportional to the length of the boundary and
therefore negligible in the thermodynamic limit, when compared to
terms proportional to the surface area.

The structure of the domain boundary is quite interesting 
as well. Two clockwise steps running under angles $\pi/2+\phi_0$ and 
$\pi/2-\phi_0$ cannot be simply joined together, because this will 
cause a mismatch in the reconstruction order. Such joints
must be accompanied by a domain boundary. 

As we saw already in the previous Section the 
model predicts that the network is unstable, suggesting that
the pattern of steps observed in STM experiments should
decay, after a sufficiently long time, into the combination of 
surface orientations depicted in Fig.\ \ref{FIG09}(b)
In the present case this time may be extremely long.
In fact once the network has been formed, the process of disentangling
it into stable orientations as the ones shown in Fig.\ \ref{FIG09}(b),
may require the investment of a large amount of free energy
to go through very unfavorable states and it may be difficult
to observe it experimentally without a careful long annealing
of the surface.

\begin{figure}
\centerline{\epsfysize=7cm \epsfbox{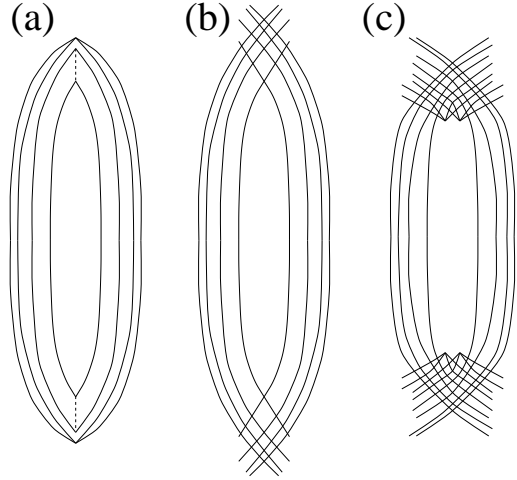}}
\caption{Arrangement of the steps surrounding the (110) facet for 
  $\tilde{\epsilon} > 0$ in the stable phase (a), in the metastable 
  phase (b) and for $\tilde{\epsilon} <0$ (c).}
\label{FIG10}
\end{figure}

Thermodynamically the metastability encountered here is highly
unusual. In e.g. a homogeneous gas-liquid system a free energy 
that is a concave function of density always leads to instability 
due to spinodal decomposition. In our system this is impeded by
topological constraints on the steps, requiring concerted mass 
transportation over relatively large distances for the 
decomposition of a network into stable surfaces. 
Therefore even a non-convex free energy can be metastable.
For describing the surface free energy 
one can use Eq. (\ref{fremin}) again.
A typical arrangement of steps around the (110) facet 
in this situation is shown in Fig.\ \ref{FIG10}(b).
In this case the shape profile along the $y$-axis of the crystal 
for vicinal orientations is expected to be of the type $z(y) \sim 
(y - y_0)^2$, due to the term proportional to $p^2$ in the surface 
free energy. A free energy expansion with a term cubic in the
step density would produce a shape profile with an exponent 3/2,
the so-called 
Pokrovsky-Talapov exponent (see, for instance, Van Beijeren and Nolden, 1987).

\vspace{1\baselineskip}
{\boldmath{$ \tilde{\epsilon} < 0$}}
\noindent{\vspace{1\baselineskip}

When a negative free energy is associated to a crossing point the
network of steps tends to condense to maximize the density of 
crossings. On the other side, entropic repulsions (Gruber and Mullins, 1967) 
favor configurations where steps are far apart and tend to stabilize 
the network. 

The competitions between repulsions and attractions may give rise to 
the two following scenarios: 1) if attractions dominate for the whole 
range of surface orientations the network fully condenses and the stable 
orientations along the $p$ direction are the $(110)$ and $(100)$ facets, 
which would be directly connected under a sharp edge; 2) if repulsions, 
possibly of entropic type become dominant, then the surface free energy 
turns concave at higher step densities and the $(110)$ facet is connected
under an angle with the rounded regions. 

As pointed out above, the surface free energy $f(p,q)$ contains terms 
which are of cubic or higher order in the step density. A term 
cubic in $p$ with a positive coefficient could well stabilize the 
free energy at not too small values of $p$. In both scenarios the 
$(110)$ facet has  
sharp edges running roughly perpendicular to the missing-row direction.

A simple model with a negative crossing energy (Carlon et al., 1996; 
Carlon, 1996)
the staggered body-centered-solid-on-solid-model (BCSOS model), is
discussed in some details in Appendix\ II. The model reproduces both
scenarios 1) and 2), depending upon temperature.

Finally in Fig.\ \ref{FIG07} the dashed lines show the truncation of 
the equilibrium shape of the $(110)$ facet by ridges connecting the facet 
to rounded areas and Fig.\ \ref{FIG10}(c) shows the expected arrangement 
of steps around the truncated facet. 
Notice that in this case there are sharp ridges between 
rounded regions covered by networks of steps and regions covered by non-crossing 
step arrays.

\vspace{2\baselineskip}
\noindent{\bf
CONCLUSION
}
\vspace{1\baselineskip}

In the present article we introduced a simple model which describes
networks of crossing steps, that have been observed in STM 
experiments on $Au(110)$ and $Pt(110)$ surfaces. The model shows 
that these networks are unstable (or at least metastable) with 
respect to faceting. 

{F}rom the calculation of the surface free energy for orientations close
to the $(110)$ facet the equilibrium shape for this facet has been
derived. For repulsive interactions between two crossing steps
($\tilde\epsilon > 0$) the facet shape is rather unusual: it is cusped
and elongated along the $[1 \bar{1} 0]$ direction. The cusps are
connected to ridges separating rounded regions.
If the interaction is attractive ($\tilde \epsilon < 0$) the $(110)$ 
facet is still elongated
with sharp edges
roughly perpendicular to the missing-row direction.
In this case there are much weaker cusps, between these edges and the smooth
facet boundaries roughly parallel to the missing--row direction.

The predicted shapes could be observed in experiments on equilibrium 
shapes of crystals with $(110)$ missing--row reconstructed  facets.

Heyraud and M\'etois (1980) studied shapes of small gold
crystals in thermal equilibrium with their vapor; in their samples 
only the $(111)$ and $(100)$ facets were observed, since the range 
of temperatures investigated ($T \approx 1000^\circ C$) is above the 
roughening temperature of the $(110)$ facet. 
To observe some of the shapes described in this article,
temperatures below the roughening and the deconstruction 
temperatures of the $(110)$ facet should be considered.

We believe that the problems connected to the metastability/instability
of the networks of steps deserve further experimental 
investigation, as well. If $\tilde\epsilon$ is positive, as we think
should be the case for gold and platinum crystals,
one should be able, starting from the metastable network of steps,
to observe a nucleation of arrays of parallel steps connected under a ridge.
Probably this could be observed in practice, e.g. in STM experiments,  
only for sufficiently high temperatures and in very pure samples.\\

\noindent{\bf Acknowledgements:}

We had several stimulating discussions with Joost Frenken, Laurens Kuipers and 
Misha Hoogeman, who drew our attention to the networks of steps discussed here in 
the first place.
E.C. aknowledges financial support from grant No.\ CHBGCT940734 from 
the European Capital
and Mobility Programme.

\vspace{2\baselineskip}
\noindent{\bf APPENDIX I: CALCULATION OF THE STEP FREE ENERGY}
\vspace{\baselineskip}

For the calculation of the step free energy we consider first the 
partition functions $G_x$ and $G_y$ of straight segments of steps 
along the $x$ and $y$ direction. The segments have energies per unit 
length equal to $\delta_x$ and $\delta_y$. 
In Fourier space one has:
\noindent
\begin{eqnarray}
G_x(k_x)& = &\sum_{n=1}^{+ \infty}  e^{-2 n 
\beta \delta_x} 
\left( e^{i n k_x} + e^{-i n k_x} \right) =
\frac{1}{e^{2 
\beta \delta_x + i k_x} - 1} +
\frac{1}{e^{2 
\beta \delta_x - i k_x} - 1}\\
G_y(k_y)& = &\sum_{n=1}^{+ \infty}  e^{- n 
\beta \delta_
y/\sqrt{2}} 
 e^{-i n k_y} = \frac{1}{e^{
\beta \delta_
y/\sqrt{2} + i k_y} - 1}
\end{eqnarray}
Notice that segments of steps along the negative $y$ direction
are not allowed.
%%EE ADDED FOR CLARITY (FROM MY THESIS)
We recall thatthe factors $\sqrt{2}$ and $2$ are the minimal length
of step segments along the $x$ and $y$ directions.
%%EE
The step is generated by all possible combinations of horizontal
and vertical segments:
\begin{eqnarray}
G(\vec k) = G_x + G_y + G_x G_y + G_y G_x + \ldots = 
\frac{G_x + G_y + 2 G_x G_y}{1 - G_x G_y}
\end{eqnarray}
The partition function of a step of lengths $L_x$ and $L_y$ is
obtained by Fourier transform:
\begin{eqnarray}
Z(L,\phi) &=& \int_{-\pi}^{+\pi}
\frac{d \vec k}{4 \pi^2} e^{i(k_y \sqrt{2} L_y + k_x L_x/2)}
G(\vec k)
\end{eqnarray}
with $L = \sqrt{L_x^2 + L_y^2}$ and $\phi = \arctan(L_x/L_y)$.
The integral 
over $k_y$ can be rewritten as a contour integral in the
variable $z = e^{i k_y}$; the integrand has a simple pole in:
\begin{eqnarray}
z &=& e^{-
\beta \delta_y/\sqrt 2} \left(1 + G_x(k_x) \right)
\end{eqnarray}
The partition function now becomes:
\begin{eqnarray}
Z(L,\phi) &=& \int_{-\pi}^{+\pi} \frac{d k_
x}{2 \pi} \;\;
e^{-L \beta f(k_x,\phi)} \ldots
\label{form1}
\end{eqnarray}
with:
\begin{eqnarray}
\beta
f(k_x,\phi) &=& \frac{i k_x}{2} \sin \phi 
%%EE \nonumber \\
%%EE&& 
%%EE ALL IN ONE LINE
+ \left[
\beta \delta_y - \sqrt{2} \log\left(1 + G_x(k_x) \right)
\right] \cos \phi
\end{eqnarray}
The dots in (\ref{form1}) denote terms which are not relevant in the
thermodynamic limit $L \to \infty$.
Using the saddle point approximation, one can evaluate (\ref{form1})
as:
\begin{eqnarray}
Z(L,\phi) &=& e^{-L \beta f_s(\phi)}
\end{eqnarray}
where the step free energy is $f_s(\phi) = f(i \bar{k}, \phi)$ and
$i \bar{k}$ the stationary point of $f(k_x,\phi)$.
Working this out one 
obtains Eq.\ \ref{singst}.

\vspace{2\baselineskip}

\noindent{\bf APPENDIX II: THE STAGGERED BCSOS MODEL}
\vspace{\baselineskip}

Without going into details we present here some relevant 
results concerning the staggered BCSOS model, which, for certain 
values of the energy parameters describes networks of crossing
steps with a {\em negative} crossing energy (for more details the reader 
may consult (Carlon et al., 1996; Carlon and Van Beijeren, 1996 II)). 

\begin{figure}
\centerline{\epsfxsize=7cm \rotate[r]{\epsfbox{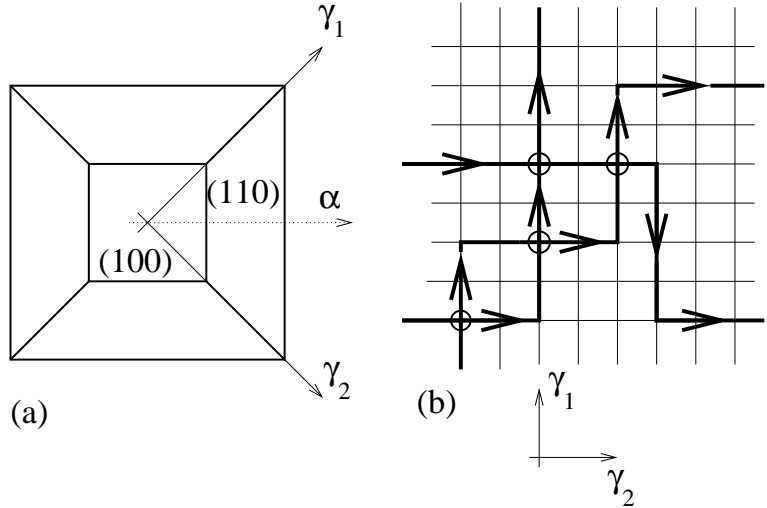}}}
\caption{(a) At $T=0$ the equilibrium shape of the BCSOS model is a
  truncated pyramid. (b) Steps are composed of elongated segments with
  few corners since the energy of a corner is much larger then the
  energy of a straight segment of step $e_K \gg e$. Crossings of steps,
  indicated by circles in the figure, are energetically favored, i.e.
  at each crossing there is a gain of energy of $2 e$.}
\label{FIG11}
\end{figure}

The model is applicable to surfaces of ionic crystals of bcc type, as for
instance CsCl, and describes all surface orientations $(ts1)$ with
$|t|+|s| \leq 1$. 
It does not have the characteristic feature, responsible for the formation
of networks in the present paper, that an up step followed by a down step
changes a reconstruction order; instead the networks are induced by a large
corner free energy, leading to steps with long segments in the principal lattice 
directions, combined with a negative value of $\tilde{\epsilon}$, which
favors crossings of steps. However, the model does illustrate the two different
scenario's for step condensation discussed 
%%EE in Sec.\ \ref{3 b}.  
%%EE THERE IS NO SECTION NUMBERING ANYMORE
above for $\tilde{\epsilon} < 0$.
%%EE

At $T=0$ the crystal has 
the shape of a truncated piramyd,
with a top $(100)$ facet and four side facets of $(110)$ type, 
as shown in Fig.\ \ref{FIG11}(a).
A step on the $(100)$ facet has an energy per unit 
of length $e > 0$ and a kink energy $e_{K} \gg e$: 
steps consist, especially at low temperatures, of elongated straight 
segments with few kinks. These segments are oriented parallel to
the two axes $\gamma_1$ and $\gamma_2$ shown in Fig.\ \ref{FIG11}(b):
differently
from $(110)$ surfaces of fcc metals these two directions are
equivalent.
Steps can cross each other, with a gain of energy 
of $2 e$ at each crossing point. In this model, thus, the crossing
energy is $\tilde{\epsilon} = -2 e < 0$.

The surface free energy has been calculated as function of the slope
parameters $p,q$ in mean--field approximation, for all possible values
of $p$ and $q$, also beyond the vicinal orientations.
To give an example,
we restrict ourselves to considering only $f(p)$, i.e. the surface free
energy for surface orientations of the type $(1p0)$, with $0 \leq p 
\leq 1$; these are all the surface orientations along the axis $\alpha$
of Fig.\ \ref{FIG11}(a).
The mean--field calculation gives a surface free energy that at low 
temperatures is not stable (Fig.\ \ref{FIG12}(a)):
step attractions, due to the energy 
gained at crossings, dominate in the whole range of surface orientations 
and the network condenses until the optimal density of steps is reached,
which in the present case corresponds to the $(110)$ facet.
In the equilibrium shape the two facets $(100)$ and $(110)$ 
touch each other at a sharp edge. 
The free energy instability is of the same form as that predicted 
in Eq.(8), with a negative value of $\tilde\epsilon$.

\begin{figure}
\centerline{\epsfxsize=7cm \rotate[r]{\epsfbox{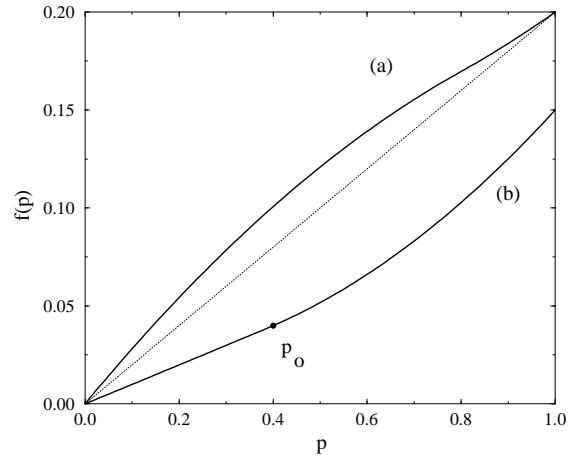}}}
\caption{Surface free energy along the axis $\alpha$ of Fig.\ 
 \protect{\ref{FIG11}}. (a) At low temperatures the free energy is not
 stable for the whole range of orientations $0 \leq p \leq 1$; the 
%%EE dashed
 dotted
 line denote the Maxwell construction. (b) At higher temperatures the range
 of orientations $p_0 \approx 0.4 \leq p \leq 1$ is now stable.}
\label{FIG12}
\end{figure}

We stress that the free energy $f(p)$ of Fig.\ \ref{FIG12}(a)
becomes convex for
sufficiently large values of $p$. This change of sign of the
second derivative of $f(p)$ corresponds to a transition from a
reconstructed to a non-reconstructed state. 
In the present case this critical point is of 
limited relevance since it occurs in an
unstable region, and there seem to be no topological restrictions impeding
the decay of an unstable orientation into two facets.

The same mean--field analysis at 
somewhat higher temperatures gives a surface
free energy of the type shown in Fig.\ \ref{FIG12}(b),
which is also concave at low $p$,
but convex at higher values of $p$. At low step densities, attractions
still dominate, yielding a concave (i.e. unstable) free energy. 
Entropic repulsions 
%%EE \cite{GruberMullins}, 
%%EE CORRECT REFERENCE
(Gruber and Mullins, 1967)
%%EE
which become important at high temperatures and high step densities, 
stabilize the network.
%%EE The dotted line in the figure is the Maxwell construction that 
%%EE connects the coexisting surface orientations: these are now the $(100)$ 
%%EE facet and a non-faceted orientation. 
The Maxwell construction gives as coexisting surface orientations the $(100)$
facet and a non-faceted orientation of slope $p_0$.
%%EE
These touch again at a sharp edge.

\vspace{2\baselineskip}
\noindent{\bf REFERENCES}

\list
  {\relax}{\setlength{\labelsep}{0em}
        \setlength{\itemindent}{-\parindent}
        \setlength{\leftmargin}{\parindent}}
    \def\newblock{\hskip .11em plus .33em minus .07em}
    \sloppy\clubpenalty4000\widowpenalty4000
    \sfcode`\.=1000\relax%

\small
\setlength{\parskip}{-4pt}

\item{} Bernasconi, M.\ and Tosatti, E., 1993,
{\it Surf. Sci. Repts.} 17:363.
\item{} Carlon, E., 1996, {\it Ph.D. Thesis}, 
Utrecht.
\item{} Carlon, E., Van Beijeren, H.\ and Mazzeo, G., 1996
{\it Phys. Rev. E} 53:R5549; 
\item{} Carlon, E.\ and Van Beijeren, H., 1996, {\it Phys. Rev. Lett.}
76:4191.
\item{} Carlon, E.\ and Van Beijeren, H., 1996, {\it preprint}.
\item{} Gimzewski, J.K., Berndt, R.\ and
Schittler, R.R., 1992, {\it Phys. Rev. B} 45:6844.
\item{} Gritsch, T., Coulman, D., Behm, R.J.\ and
Ertl, G,, 1991, {\it Surf. Sci.} 257:297. 
\item{} Gruber, E.E.\ and Mullins, W.W., 1967,
{\it J. Phys. Chem. Solids} 28:875.
\item{} Heyraud, J.C.\ and M\'etois, J.J., 1980, {\it Acta Metall.}
28:1789.
\item{} Kuipers, L., 1994, {\it Ph.D. Thesis} Amsterdam.
\item{} Jayaprakash, C., Rottman, C. and Saam, W.F., 1984, {\it Phys. Rev. B}
30:6549.
\item{} Van Beijeren, H.\ and Nolden, I., 1987, in: {\it 
Structure and Dynamics of Surfaces, Vol.\ 2}, 
W.\ Schommers and P.\ von Blanckenhagen eds., Springer-Verlag,
Berlin, 1987.

\end{document}